\documentclass[preprint,aps,tightenlines,floatfix,showpacs]{revtex4}

\usepackage{graphics}
\usepackage{bm}
\usepackage{amsmath}
\usepackage{amssymb}
\begin{document}

\title
{Simple function forms and nucleon-nucleus total cross sections}
\author{P. K. Deb}
\email{pdeb@mps.ohio-state.edu}
\affiliation
{Department of Physics, The Ohio State University, Columbus, OH 43210, U.S.A.}
\author{K. Amos}
\email{amos@physics.unimelb.edu.au}
\affiliation
{School of Physics, The University of Melbourne, Victoria 3010, Australia}

\date{\today}
\begin{abstract}
Total cross sections for neutron scattering with energies between 10 and 600 
MeV and from nine nuclei spanning the mass range from ${}^6$Li to ${}^{238}$U 
have been analyzed using a simple function of three parameters. The values of 
those parameters with which neutron total cross-section data are replicated 
vary smoothly with energy and target mass and may themselves be represented 
by functions of energy and mass.
\end{abstract}
\maketitle

\section{Introduction}

Total cross sections from the scattering of neutrons by nuclei are required in
a number of fields of study which range over problems in basic science as well
as many of applied nature. It would be utilitarian if such scattering data were
well approximated by a simple convenient function form with which predictions 
could be made for cases of energies and/or masses as yet to be measured.  
Recently it has been shown~\cite{Am02,De03} that such forms may exist for 
proton total reaction cross sections.  Herein we consider that concept further 
to reproduce the measured total cross sections from neutron scattering for 
energies to 600 MeV and from nine nuclei ranging in mass between ${}^6$Li and 
${}^{238}$U. These suffice to show that such forms will also be applicable in 
dealing with other stable nuclei since their neutron total cross sections vary
so similarly with energy~\cite{Ko03}.

Total scattering cross sections for neutrons from nuclei have been well 
reproduced by using optical potentials. In particular, the data (to 300 MeV) 
from the same nine nuclei we consider, compare quite well with predictions made
using a $g$-folding method to form nonlocal optical potentials~\cite{Amos02}, 
though there are some notable discrepancies. Alternatively, in a recent study 
Koning and Delaroche~\cite{Ko03} gave a detailed specification of 
phenomenological global optical model potentials determined by fits to quite a 
vast amount of data, and in particular to the neutron total scattering cross 
sections we consider herein. However, as we show in the case of the total cross
sections from 10 to 600 MeV, there is a simple function form one can use to 
allow estimates to be made quickly without recourse to optical potential 
calculations. Furthermore we shall show that the required values of the three 
parameters of that function form themselves trend sufficiently smoothly with 
energy and mass suggesting that they too may be represented by functional 
forms.

\section{ Formalism}

The total cross sections for neutron scattering from nuclei can be expressed 
in terms of partial wave scattering ($S$) matrices specified at energies 
$E\propto k^2$, by
\begin{equation}
S^{\pm}_l \equiv S^{\pm}_l(k) = e^{2i\delta^{\pm}_l(k)} =
\eta^{\pm}_l(k)e^{2i\Re\left[ \delta^{\pm}_l(k) \right] }\ ,
\end{equation}
where $\delta^\pm_l(k)$ are the (complex) scattering phase shifts and 
$\eta^{\pm}_l(k)$ are the moduli of the $S$ matrices. The superscript 
designates $j = l\pm 1/2$. In terms of these quantities, the elastic, reaction
(absorption), and total cross sections respectively are given by
\begin{eqnarray}
\sigma_{\text{el}}(E) & = & \frac{\pi}{k^2} \sum^{\infty}_{l = 0} \left\{
\left(l + 1 \right)\left|S^+_l(k) - 1 \right|^2 + l\left|S^-_l(k) - 1\right|^2 
\right\} = \frac{\pi}{k^2} \sum_l \sigma_l^{(el)}\\
\sigma_{\text{R}}(E) & = & \frac{\pi}{k^2} \sum^{\infty}_{l = 0}\left\{ \left(
l + 1 \right) \left[ 1 - \eta^+_l(k)^2 \right] + l \left[ 1 - \eta^-_l(k)^2 
\right] \right\} = \frac{\pi}{k^2} \sum_l \sigma_l^{(R)}\ ,
\end{eqnarray}
and
\begin{eqnarray}
\sigma_{\text{TOT}}(E) & = & \sigma_{\text{el}}(E) + \sigma_{\text{R}}(E)
= \frac{\pi}{k^2} \left[\sigma_l^{(el)} + \sigma_l^{(R)}\right] 
= \frac{2\pi}{k^2} \sum_l \sigma_l^{(TOT)}\ ,
\nonumber\\ 
\sigma_l^{(TOT)} & = &  \left( l + 1 \right) 
\left\{ 1 - \eta^+_l(k)\cos\left( 2\Re\left[ \delta^+_l(k) \right] \right) 
\right\} + l\left\{1 - \eta^-_l(k) \cos\left( 2\Re\left[ \delta^-_l(k) \right] 
\right) \right\}\ .
\label{SumTOT}
\end{eqnarray}
Therein the $\sigma_l^{(X)}$ are defined as partial cross sections of the total
elastic, total reaction, and total scattering itself.   For proton scattering, 
because Coulomb amplitudes diverge at zero degree scattering, only total 
reaction cross sections are measured. Nonetheless study of such 
data~\cite{Am02,De03} established that partial total reaction cross sections 
$\sigma_l^{(R)}(E)$ may be described by the simple function form
\begin{equation}
\sigma_l^{(R)}(E) = (2l+1) \left[1 + e^{\frac{(l-l_0)}{a}}\right]^{-1} + 
\epsilon\ (2l_0 + 1)\ e^{\frac{(l-l_0)}{a}} \left[1 + e^{\frac{(l-l_0)}{a}}
\right]^{-2}\ ,
\label{Fnform}
\end{equation}
with the tabulated values of $l_0(E,A)$, $a(E,A)$, and $\epsilon(E,A)$ all 
varying smoothly with energy and mass.  Those studies were initiated with the
partial reaction cross sections determined by using complex, non-local, 
energy-dependent, optical potentials generated from a $g$-folding 
formalism~\cite{Am00}. While those $g$-folding calculations did not always give
excellent reproduction of the measured data (from $\sim$ 20 to 300 MeV for 
which one may assume that the method of analysis is credible), they did show a 
pattern for the partial reaction cross sections that suggest the simple 
function form given in Eq.~(\ref{Fnform}).  With that form excellent 
reproduction of the proton total reaction cross sections for many targets and
over a wide range of energies were found with parameter values that varied 
smoothly with energy and mass.

Herein we establish that the partial total cross sections for scattering of 
neutrons from nuclei can also be so expressed and we suggest forms, at least
first average result forms, for the characteristic energy and mass variations 
of the three parameters involved.  Nine nuclei, $^6$Li, $^{12}$C, $^{19}$F, 
$^{40}$Ca, $^{89}$Y, $^{184}$W, $^{197}$Au, $^{208}$Pb and $^{238}$U, for
which a large set of experimental data exist, are considered.  Also those 
nuclei span essentially the whole range of target mass.  However, to set up an
appropriate simple function form, initial partial total cross sections must be 
defined by some method that is physically reasonable.  Thereafter the measured
total cross-section values themselves can be used to tune details, and of the 
parameter $l_0$ in particular. We chose to use results from $g$-folding optical
potential calculations to give those starting values.

\section{ Results and discussions}

That a function form for total cross sections is feasible has been suggested
previously in dealing with energies to 300 MeV from a few nuclei~\cite{Amos02}
and by using a $g$-folding prescription for the nucleon-nucleus optical 
potentials.  At the same time, studies of the partial reaction total cross 
sections for proton scattering~\cite{Am02,De03} found that a form as given in
Eq.~(\ref{Fnform}) was most suitable. A similar form can be used to map the 
partial total cross sections given by the $g$-folding potential calculations 
and thence by suitable adjustments for their sums to give the measured total 
cross sections.  Of note is that, with increasing energy, the form of the 
simple function (Eq.~(\ref{Fnform})) can be approximated by a sharp fall at 
$l = l_0(E) = l_{max}$ giving a triangle in angular momentum space.  In that 
case, the total reaction cross section equates to the area of that triangle and
\begin{equation}
\sigma_R \Rightarrow \frac{\pi}{2 k^2} l_{max}(2l_{max} + 1)
\approx \frac{\pi}{k^2} l_{max}^2\ .
\end{equation}
Then with $l_{max} \sim kR$ at high energies, $\sigma_R \Rightarrow \pi R^2$ 
the geometric cross section as required. Furthermore, for high enough energies
then, the total cross section is double that value. This is an asymptotic 
behavior one can assume for the $l_0$ values to be used with the total cross 
sections.

The function form results we display in the following set of figures were 
obtained by starting with $g$-folding model results at energies of 10 to 100 
MeV in steps of 10 MeV, then to 350 MeV in steps of 25 MeV, and thereafter in
steps of 50 MeV to 600 MeV.  The $g$-matrices used above pion threshold were 
those obtained from an optical potential correction to the BonnB 
force~\cite{Ge98} which, while approximating the effects of resonance terms 
such as virtual excitation of the $\Delta$, may still be somewhat inadequate 
for use in nucleon-nucleus scattering above 300 MeV.  Also relativistic 
effects in scattering, other than simply the use of relativistic kinematics 
in the distorted wave approximation (DWA) approach, are to be expected.  
Nonetheless the DWA results are used only to find a sensible starting set of 
the function form parameters $l_0, a$, and $\epsilon$ from which to find ones 
that reproduce the measured total cross-section data. One must also note that 
the $g$-folding potentials for most of the nuclei considered were formed using 
extremely simple model prescriptions of their ground states. A previous 
study~\cite{Amos02} revealed that with good spectroscopy the $g$-folding 
approach gives much better results in comparison with data than that approach 
did when simple packed shell prescriptions for the structure of targets were 
used. That was also the case when scattering from exotic, so-called nucleon 
halo, nuclei were studied~\cite{Am00,La01}. 

The results from analyses of $40A$ MeV scattering of $^6$He ion from hydrogen 
targets~\cite{La01} lead to a note of caution for the use of the trends we set
out here. Our results are for a range of energies and for a diverse set of 
{\em stable} nuclear targets. Total cross sections with unstable halo nuclei 
may be considerably larger than one expects if they were assumed adequately 
described by standard shell model wave functions. Indeed at $40A$ MeV the total
reaction cross section for $^6$He-hydrogen scattering was 16-17\% larger than 
found using the standard shell model prescription. That and the momentum 
transfer properties of the ${}^6$He-p differential cross section were 
convincing evidence of the neutron-halo 
nature of $^6$He.  We proceed then with the caveat that specific structure 
properties may be needed as variation to the functional forms we deduce. But, 
given the results found with the diverse (nine) nuclei considered, we believe 
that such would need be very significant structure aspects, such as a halo, to
be of import.

\begin{figure}
\centering
\scalebox{0.7}{\includegraphics{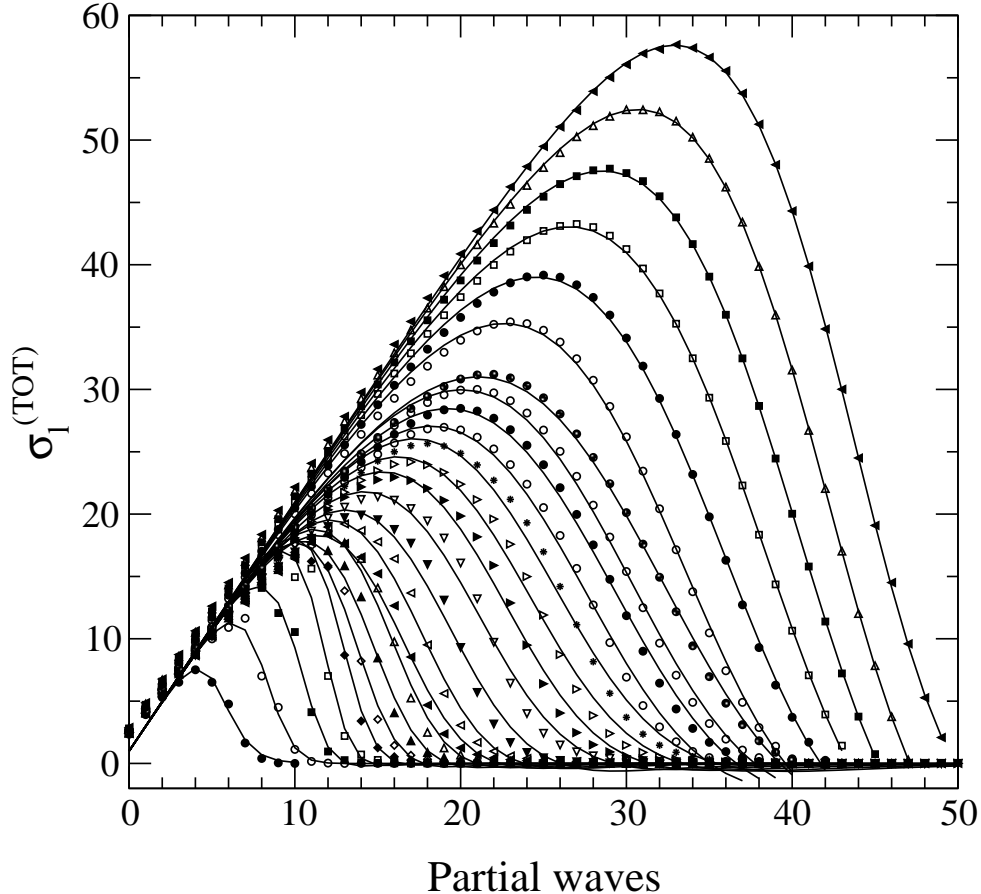}}
\caption{\label{Pb208-partials} 
The partial total cross sections for scattering of neutrons from ${}^{208}$Pb
with the set of energies between 10 and 600 MeV specified in the text. The 
largest energy has the broadest spread of values.}
\end{figure}

\begin{figure}
\centering
\scalebox{0.8}{\includegraphics{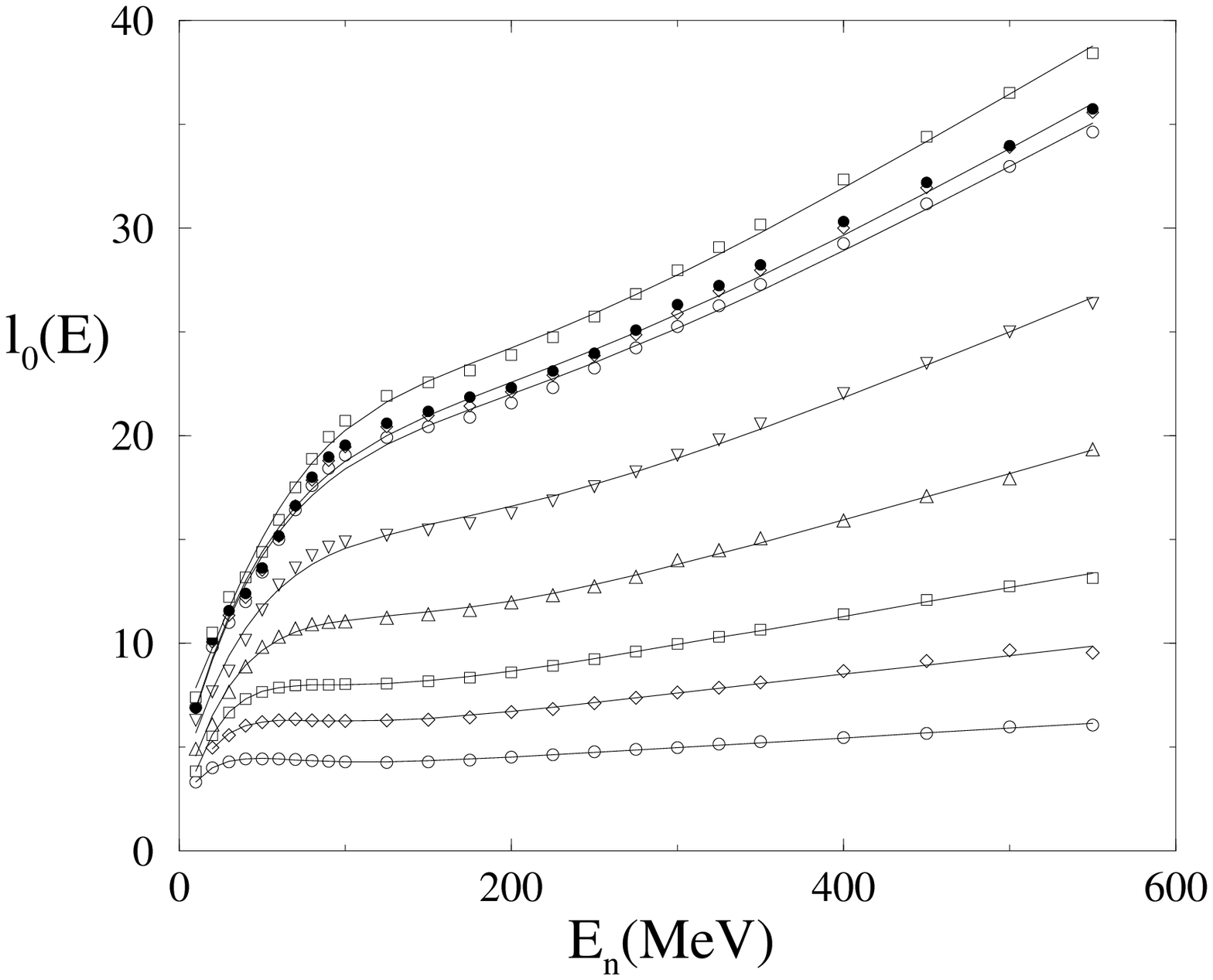}}
\caption{\label{l0vsE} 
The values of $l_0$ that fit neutron total scattering cross-section data from 
the nine nuclei considered and for energies between 10 and 600 MeV. The curves
portray the best fits found by taking a function form for $l_0(E)$.}
\end{figure}

\begin{figure}
\centering
\scalebox{0.8}{\includegraphics*{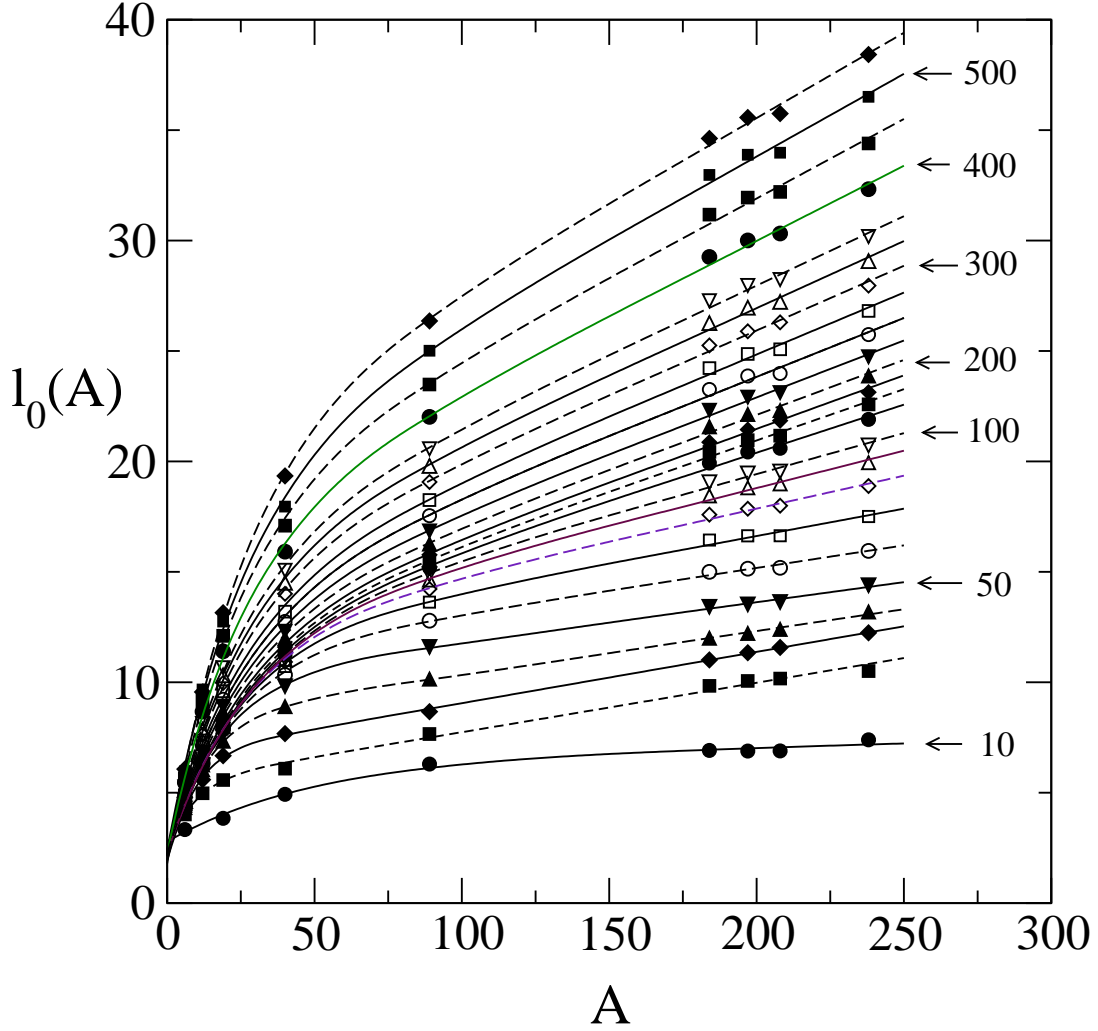}}
\caption{\label{l0vsA} 
The values of $l_0$ depicted in Fig.~\ref{l0vsE} as they vary with mass for 
all of the energies considered. Some of those energies are indicated on the 
diagram and the curves are splines linking best fit values for each mass 
assuming a function form for $l_0(A)$.}
\end{figure}

While we have used the partial total cross sections from DWA results for 
neutron scattering from all the nine nuclei chosen and at all of the energies
indicated, only those obtained for ${}^{208}$Pb are shown in 
Fig.~\ref{Pb208-partials}. The results from calculations of scattering from 
the other eight nuclei have similar form.  The `data' shown as diverse open 
and closed symbols in Fig.~\ref{Pb208-partials} are the specific values found 
from the $g$-folding optical model calculations. Each curve shown therein is 
the result of a search for the best fit values of the three parameters, $l_0$,
$a$, and $\epsilon$ that map Eq.~(\ref{Fnform}) (now for total neutron cross 
sections) to these `data'.
From the sets of values that result from that fitting process, the two 
parameters $a$ and $\epsilon$ can themselves be expressed by the parabolic 
functions 
\begin{eqnarray}
a\ &=& {\phantom{-}}1.29\ +\ 0.00250\ E\ -\ 1.76\ \times 10^{-6}\ E^2\ , 
\nonumber\\
\epsilon\ &=& -1.47\ -\ 0.00234\ E\ +\ 4.16\ \times 10^{-6}\ E^2\ ,
\label{Eps}
\end{eqnarray}
where the target energy E is in MeV. There was no conclusive evidence for a 
mass variation of them. With $a$ and $\epsilon$ so fixed, we then adjusted the
values of $l_0$ in each case so that actual measured neutron total 
cross-section data were fit using Eq.~(\ref{Fnform}).  Numerical values for
$l_0$ from that process are presented in Table~\ref{l0-table}.
\begin{table}
\caption{\label{l0-table}
$l0$ values with which the  function form Eq.~(\ref{Fnform}) fits neutron 
total cross sections.}
\begin{ruledtabular}
\begin{tabular}{cccccccccc}
E (MeV) & $^6$Li & $^{12}$C & $^{19}$F & $^{40}$Ca & $^{89}$Y & $^{184}$W &
$^{197}$Au & $^{208}$Pb & $^{238}$U  \\
\hline
10 & 3.330 & 3.650 & 3.838 & 4.925 & 6.297 & 6.916 & 6.883 & 6.892 & 7.397\\
20 & 4.016 & 4.974 & 5.573 & 6.088 & 7.657 & 9.838 & 10.059 & 10.169 & 10.508 \\
30 & 4.292 & 5.589 & 6.677 & 7.675 & 8.671 & 11.013 & 11.337 & 11.578 & 12.241\\
40 & 4.432 & 6.039 & 7.329 & 8.898 & 10.141 & 11.993 & 12.212 & 12.393 & 13.184\\
50 & 4.447 & 6.200 & 7.672 & 9.822 & 11.602 & 13.418 & 13.526 & 13.635 & 14.392\\
60 & 4.435 & 6.296 & 7.873 & 10.331 & 12.791 & 15.001 & 15.143 & 15.181 & 15.950\\
70 & 4.404 & 6.348 & 7.979 & 10.718 & 13.629 & 16.439 & 16.632 & 16.634 & 17.506\\
80 & 4.353 & 6.305 & 8.000 & 10.922 & 14.221 & 17.591 & 17.857 & 17.996 & 18.884\\
90 & 4.324 & 6.255 & 8.003 & 11.036 & 14.631 & 18.438 & 18.808 & 18.982 & 19.940\\
100 & 4.292 & 6.259 & 8.040 & 11.071 & 14.891 & 19.058 & 19.459 & 19.541 & 20.726\\
125 & 4.261 & 6.284 & 8.067 & 11.241 & 15.190 & 19.924 & 20.427 & 20.596 & 21.900\\
150 & 4.303 & 6.315 & 8.189 & 11.404 & 15.461 & 20.432 & 20.960 & 21.167 & 22.584\\
175 & 4.387 & 6.436 & 8.362 & 11.597 & 15.771 & 20.871 & 21.441 & 21.843 & 23.129\\
200 & 4.515 & 6.686 & 8.610 & 11.981 & 16.256 & 21.567 & 22.125 & 22.304 & 23.870\\
225 & 4.648 & 6.847 & 8.921 & 12.307 & 16.850 & 22.313 & 22.910 & 23.112 & 24.735\\
250 & 4.767 & 7.113 & 9.226 & 12.756 & 17.543 & 23.255 & 23.866 & 23.981 & 25.745\\
275 & 4.883 & 7.369 & 9.593 & 13.196 & 18.250 & 24.226 & 24.866 & 25.076 & 26.814\\
300 & 4.974 & 7.621 & 9.967 & 14.008 & 19.071 & 25.249 & 25.894 & 26.297 & 27.961\\
325 & 5.143 & 7.850 & 10.312 & 14.501 & 19.794 & 26.262 & 26.962 & 27.221 & 29.069\\
350 & 5.265 & 8.131 & 10.658 & 15.069 & 20.569 & 27.277 & 27.966 & 28.236 & 30.180\\
400 & 5.456 & 8.677 & 11.399 & 15.915 & 22.015 & 29.255 & 30.007 & 30.319 & 32.327\\
450 & 5.656 & 9.159 & 12.102 & 17.091 & 23.482 & 31.173 & 31.946 & 32.202 & 34.398\\
500 & 5.966 & 9.674 & 12.751 & 17.953 & 25.011 & 32.971 & 33.887 & 33.978 & 36.510\\
550 & 6.069 & 9.559 & 13.146 & 19.341 & 26.362 & 34.624 & 35.574 & 35.749 & 38.425\\
\end{tabular}
\end{ruledtabular}
\end{table}
The values of  $l_0$ increase monotonically with both mass and energy and that
is most evident in Fig.~\ref{l0vsE} where the optimal values $l_0(E)$ are 
presented as diverse filled or open symbols. The set for each of the  masses 
(from 6 to 238) are given by those that increase in value respectively at 600
MeV. While that is obvious for most cases, note that there is some degree of
overlap in the values for ${}^{197}$Au (opaque diamonds) and for ${}^{208}$Pb
(filled circles). The curves are the shapes deduced by a function of energy 
for the $l_0(E)$ that will be discussed subsequently. 
Plotting the values of $l_0$ against mass also reveals smooth trends as is 
evident in Fig.~\ref{l0vsA}. Some actual energies are indicated by the numbers
shown in this diagram.  Again the curves shown in the figure are the results 
found on taking a functional form for $l_0(A)$ at each energy, and that too 
will be discussed later. 

\begin{figure}
\centering \scalebox{0.8}{\includegraphics{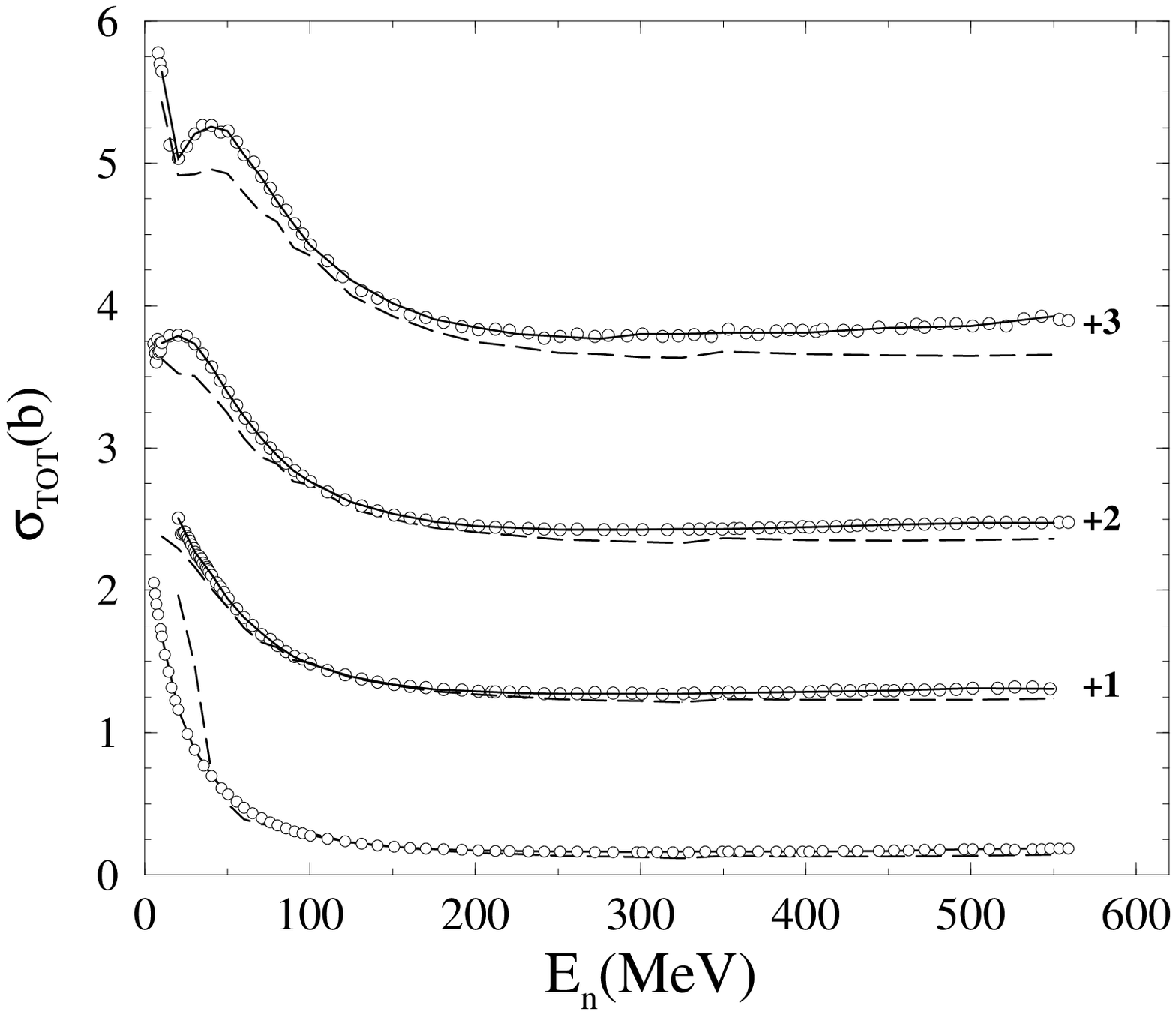}}
\caption{\label{6-40-nTotX}
Total cross sections for neutrons scattered from $^6$Li, $^{12}$C, $^{19}$F, 
and $^{40}$Ca.  The results have been scaled as described in the text to 
provide clarity.}
\end{figure}

\begin{figure}
\centering \scalebox{0.8}{\includegraphics{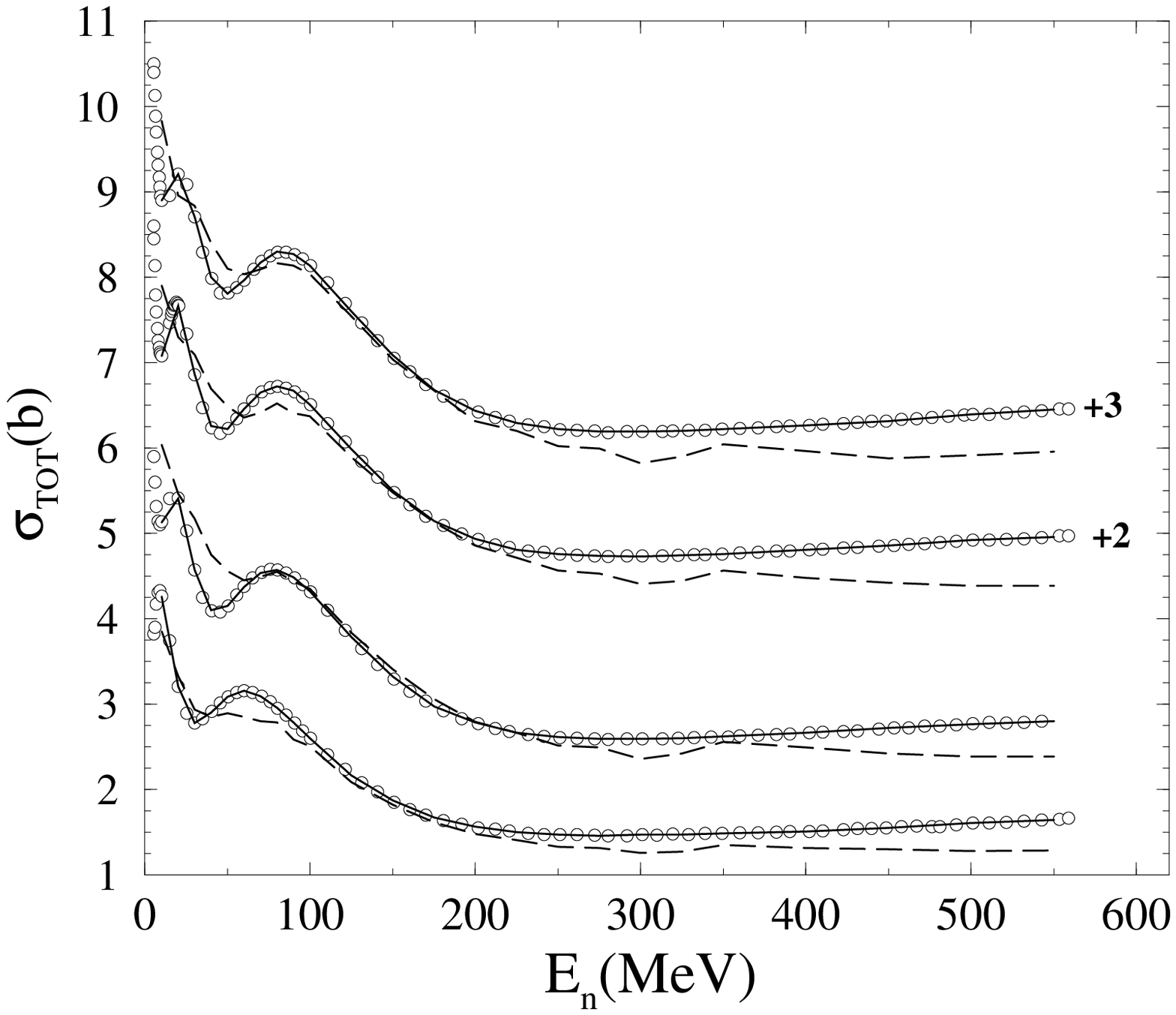}}
\caption{\label{89-238-nTotX}
Total cross sections for neutrons scattered from $^{89}$Y, $^{184}$W, 
$^{197}$Au, and $^{238}$U.  The results have been scaled as described in the 
text to provide clarity.}
\end{figure}

\begin{figure}
\centering \scalebox{0.8}{\includegraphics{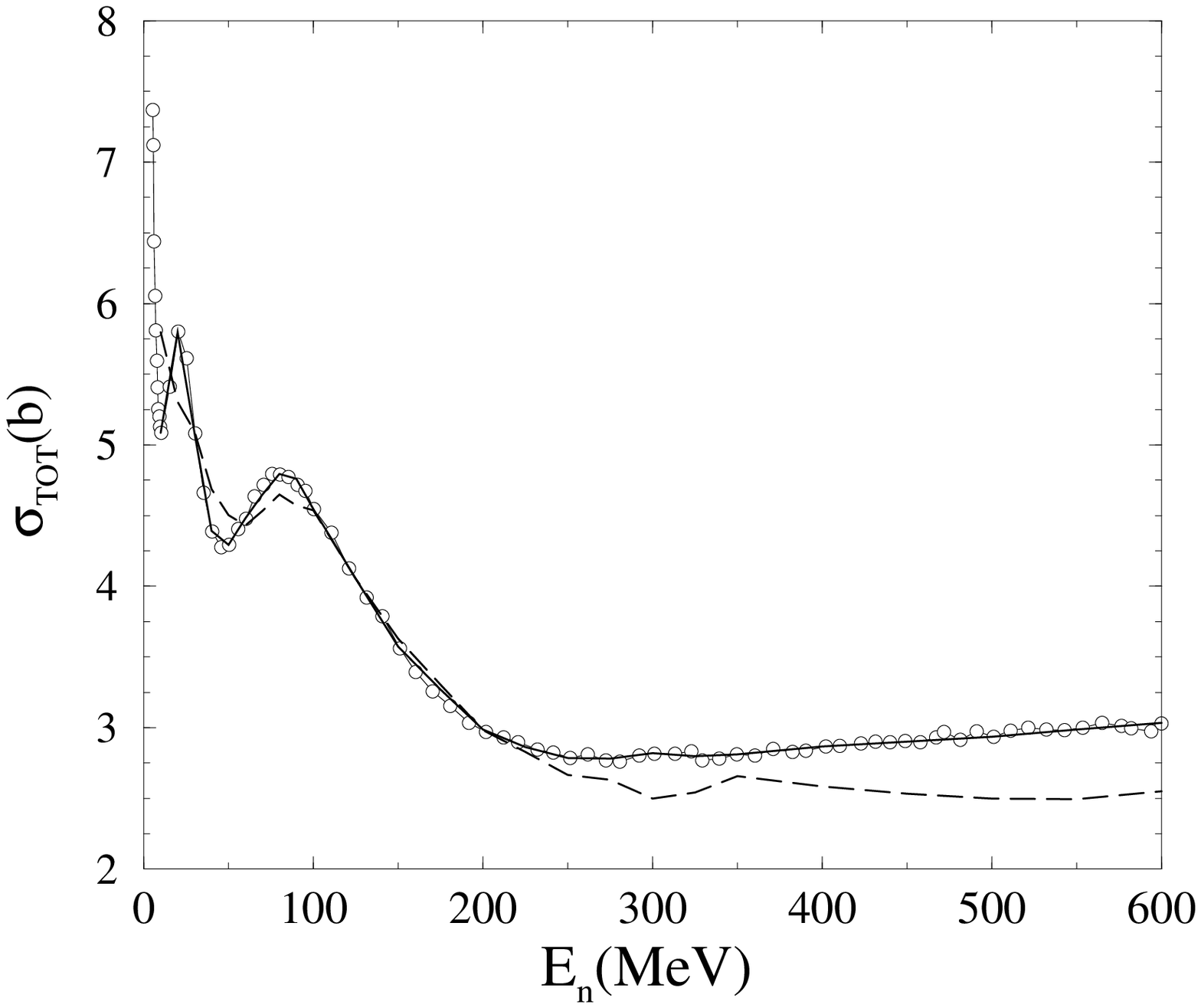}}
\caption{\label{208-nTotX}
Total cross section for neutrons scattered from  $^{208}$Pb.}
\end{figure}

The total neutron scattering cross sections generated using the function form
for partial total cross sections with the tabled values of $l_0$
and the energy function forms of Eq.~(\ref{Eps}) for $a$ and $\epsilon$,
are shown in Figs.~\ref{6-40-nTotX}, \ref{89-238-nTotX}, and \ref{208-nTotX}.  
They are displayed by the continuous lines that closely match the data which 
are portrayed by opaque circles. The data that was taken from a survey by 
Abfalterer~{\it et al.}~\cite{Abf01} which includes data measured at LANSCE 
that are supplementary and additional to those published earlier by 
Finlay~{\it et al.}~\cite{Fin93}. For comparison we show results obtained from 
calculations made using $g$-folding optical potentials~\cite{Amos02}. Dashed 
lines represent the predictions obtained from those microscopic optical 
potential calculations. Clearly for energies 300 MeV and higher, those 
predictions fail.
 
The total cross sections for neutrons scattered from the four lightest nuclei
considered are compared with data in Fig.~\ref{6-40-nTotX}.
Therein from bottom to top are shown the results for $^6$Li, $^{12}$C, 
$^{19}$F, and $^{40}$Ca with shifts of 1, 2 and 3 b made for the latter three
cases respectively to facilitate inspection of the
four sets. A slightly different scaling is used in Fig.~\ref{89-238-nTotX} in 
which the total neutron scattering cross sections from the nuclei ${}^{89}$Y 
(unscaled), ${}^{184}$W (unscaled), ${}^{197}$Au (shifted by 2 b), and 
${}^{238}$U (shifted by 3 b) are compared with the base $g$-folding optical 
potential results and with the function forms with the optimal parameters. 
Again 
the $g$-folding potential results are displayed by the dashed curves while 
those of the function form are shown by the solid curves.
Finally we show in Fig.~\ref{208-nTotX}, the results for neutron scattering 
from ${}^{208}$Pb. In this case we used Skyrme-Hartree-Fock model (SKM*) 
densities~\cite{Br00} to form the $g$-folding optical potentials. That 
structure when used to analyze proton and neutron scattering differential 
cross sections at 65 and 200 MeV gave quite excellent results~\cite{Ka02}.  
Indeed those analyzes were able to show selectivity for that SKM* model of 
structure and for the neutron skin thickness of 0.17 fm that it proposed. 
Using the SKM* model structure, the $g$-folding optical potentials gave the 
total cross sections shown by the dashed curve in Fig.~\ref{208-nTotX}. Of 
all the results, we believe these for ${}^{208}$Pb point most strongly to a 
need to improve on the $g$-folding prescription as is used currently when 
energies are at and above pion threshold.  Nonetheless, it does do quite well
for lower energies, most notably giving a reasonable account of the 
Ramsauer resonances~\cite{Ko03} below 100 MeV. However, as with the other 
results, these $g$-folding values serve only to define a set of partial 
cross sections from which an initial guess at the parameter values of the 
function form is specified. With adjustment that form produces the solid curve
shown in Fig.~\ref{208-nTotX} which is an excellent reproduction of the data,
as it was designed to do.  But the key feature is that the 
optimal fit parameter values still vary smoothly with mass and energy.

Without seeking further functional properties of the parameters, one could 
proceed as we have done this far but by using many more cases of target mass
and scattering energies so that a parameter tabulation as a data base may be 
formed with which any required value of total scattering cross section might 
be reasonably predicted (i.e. to within a few percent) by suitable 
interpolation on the data base, and the result used in Eq.~(\ref{Fnform}).

\subsection{The parameters as functions of energy}

As noted previously, the two parameters $a$ and $\epsilon$ can be chosen to 
have the parabolic forms in energy as given by Eq.~(\ref{Eps}). Once they are 
set the required values of $l_0(E,A)$ vary smoothly and monotonically with 
both $E$ and $A$ in giving the  partial cross-section sums that perfectly 
match measured values of the total cross sections.

For energies above 250 MeV, the $l_0$ values approximate well as straight 
lines and a likely representation of all of the sets of $l_0$ values is found 
with the energy dependent function, 
\begin{equation}
l_0^{\rm th}(E) = c_1 E\ +\ c_2\ -\ c_3 \left[1 - \frac{E}{E_0} \right]\ 
e^{-\beta E}\ .
\label{fnofE}
\end{equation}
The values of the parameters that lead to the curves depicted in 
Fig.~\ref{l0vsE} are listed in Table~\ref{Fnparam}. Note that a result for 
${}^{208}$Pb is not shown in Fig.~\ref{l0vsE} to avoid confusion with that for
${}^{197}$Au which is. The result for ${}^{208}$Pb nonetheless is as good a fit
as found in the other eight cases. 
\begin{table}
\begin{ruledtabular}
\caption{\label{Fnparam}  Values of parameters defining $l_0(E)$}
\begin{tabular}{cccccccc}
A & $c_1$ & $c_2$ & $c_3$ & E$_0$ & $\beta$ & $\chi^2$ & $\chi^2 (< 100) $\\  
\hline
${}^6$Li & 4.665 $10^{-3}$ & 3.582 & 1.537 & 13.87 & 
3.670 $10^{-2}$ & 0.025 & \\
${}^{12}$C & 9.103 $10^{-3}$ & 4.865 & 3.449 & 21.35 &
3.285 $10^{-2}$ & 0.30 & \\
${}^{19}$F &  1.374 $10^{-2}$ & 5.808 & 4.794 & 24.42 &
2.880 $10^{-2}$ & 0.89 & \\
${}^{40}$Ca & 2.272 $10^{-2}$ & 6.820 & 4.896 & 25.97 &
1.937 $10^{-2}$ & 0.73 & \\
${}^{89}$Y & 3.31 $10^{-2}$ & 8.357 & 5.256 & 29.47 &
1.470 $10^{-2}$ & 2.59 & 2.0 \\
${}^{184}$W & 4.27 $10^{-2}$ & 11.50 & 7.574 & 43.73 &
1.310 $10^{-2}$ & 4.2 & 3.0 \\
${}^{197}$Au & 4.41 $10^{-2}$ & 11.65 & 7.635 & 43.96 & 
1.277 $10^{-2}$ & 4.5 & 3.2 \\
${}^{208}$Pb & 4.067 $10^{-2}$ & 13.43 & 9.402 & 62.51 & 
1.400 $10^{-2}$ & 5.0 & 3.3 \\
${}^{238}$U & 4.75 $10^{-2}$  & 12.56 & 8.081 & 46.51 & 
1.235 $10^{-2}$  & 4.1 & 2.9 \\
\end{tabular}
\end{ruledtabular} 
\end{table}
In Table~\ref{Fnparam}, 
the last two columns give values of $\chi^2$ which in this case are defined by
\begin{equation}
\chi^2 = \sum_i \left[l_0^{\rm th}(E_i) - l_0(E_i) \right]^2\ ,
\label{chisq}
\end{equation}
with the sum extending over the 24 energies used.  For the heavier masses the
values of $\chi^2$ that result when the sums are restricted to energies below
100 MeV (10  points) are given in the last column. They reveal that the 
mismatch occurs at those low energies particularly. Note however, the function
for the parameter variation was chosen solely by inspection. No particular 
physical constraint was sought and so alternate function forms are not 
excluded.  This is one reason why we have not proceeded further and sought a 
mass dependence in the coefficients $c_1, c_2, c_3, E_0, \beta$ themselves.

\begin{figure}
\centering \scalebox{0.8}{\includegraphics{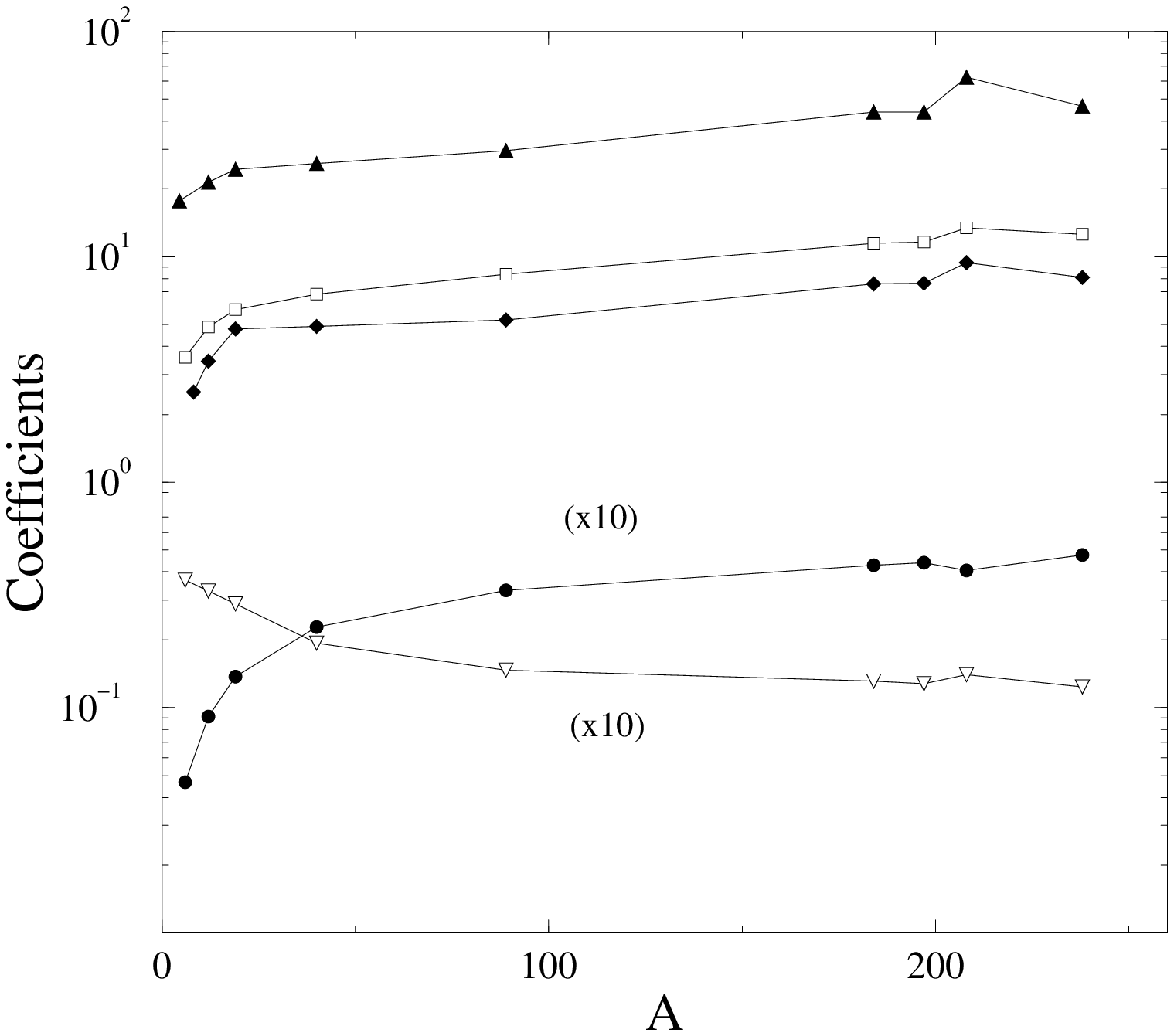}}
\caption{\label{Afit-params}
The coefficients of $l_0^{\rm th}(E)$ for each nucleus. The separate results 
are identified in the text.}
\end{figure}

Of the parameter values for the  $l_0^{\rm th}(E)$, those for ${}^{208}$Pb 
differ most from smooth progressions in mass as is evident in 
Fig.~\ref{Afit-params}.  
Therein the values of the parameters defining $l_0^{\rm th}(E)$ are plotted 
with the connecting lines simply to guide the eye. The values for $c_1$ (filled
circles) and of $\beta$ (open down triangles) have been multiplied by 10 for 
convenience of plotting.  The other parameter values are identified as $c_2$ 
(open squares), $c_3$ (filled diamonds), and $E_0$ (filled triangles). Clearly
there is a smooth mass trend of these values with the exception of the entries
for ${}^{208}$Pb.  But the ${}^{208}$Pb values are based only on achieving the
smallest $\chi^2$ value as defined by Eq.~(\ref{chisq}). Using parameter 
values consistent with the smooth mass trend, the $\chi^2$ for the fit to the 
${}^{208}$Pb values doubles at most.

\begin{figure}
\centering \scalebox{0.8}{\includegraphics{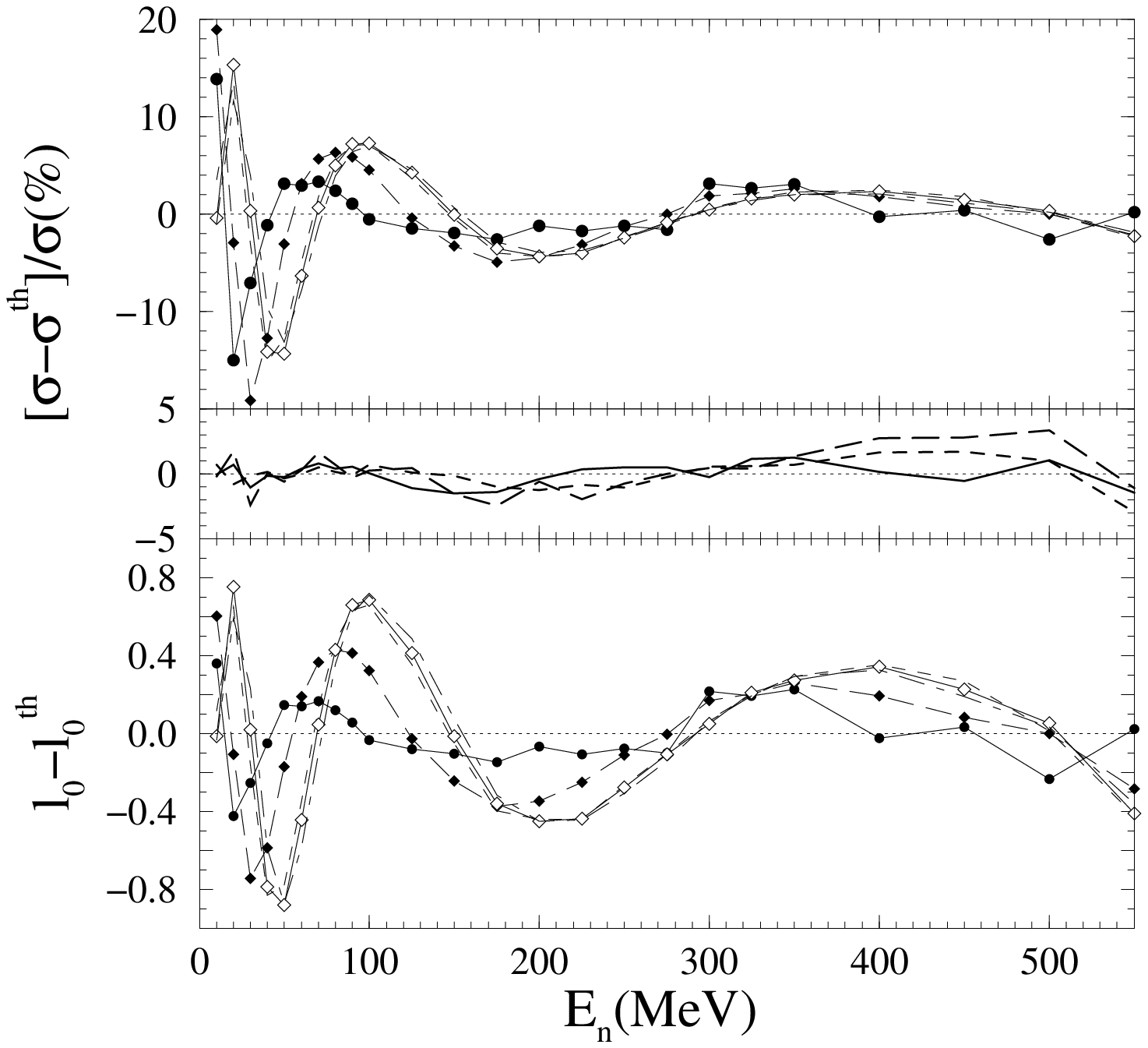}}
\caption{\label{Diffs-l0-xsecs}
The percentage differences between actual total cross section values and
those generates using the three parameter prescription with parameter values
set by the energy function forms for $l_0(E),\ a,\ {\rm and}\ \epsilon$. Those differences 
for ${}^{40}$Ca and heavier nuclei are depicted in the top segment, while those
for ${}^6$Li, ${}^{12}$C, and ${}^{19}$F are given in the middle segment.
In the bottom segment are the differences between the optimal data fit values
of $l_0(E)$ and those specified by using Eq.~(\ref{fnofE}) for ${}^{40}$Ca and
heavier nuclei.}
\end{figure}

But use of the function form of Eq.~(\ref{fnofE}) for $l_0(E)$, along with 
those of Eq.~(\ref{Eps}) for $a$ and $\epsilon$, with Eq.~(\ref{Fnform}), as 
yet do not replicate the measured total cross sections well enough at all 
energies; another reason why we do not as yet seek mass dependent forms for
the coefficients in Eq.~(\ref{fnofE}).  
We consider that an appropriate criterion is that the measured cross
sections should be replicated to within $\pm$ 5\%.  The percentage differences
in cross sections for each nucleus considered are displayed in the top two 
segments of Fig.~\ref{Diffs-l0-xsecs}. In the top segment, those differences 
for ${}^{40}$Ca and heavier nuclei are shown.  Curiously these variations
look sinusoidal with argument proportional to $E^{\frac{1}{3}}$. In the middle 
segment the differences for the three light mass nuclei are given with the 
solid, dashed and long-dashed curves depicting the 
values for ${}^6$Li, ${}^{12}$C, and ${}^{19}$F respectively.  Clearly the 
reasonable fit criterion has been met for the light masses for all energies.
That is so also for the heavier nuclei but only for energies above 100 MeV. 
There is too large a mismatch for the heavy nuclei at lower energies however.
This mismatch reflects the differences between the actual best fit values of 
$l_0(E)$ and those defined by the function form Eq.~(\ref{fnofE}), and which 
differences for just the heavy nuclei are shown in the bottom segment of 
Fig.~\ref{Diffs-l0-xsecs}.
Only the values
for ${}^{40}$Ca and heavier nuclei are shown as the differences for the light 
mass nuclei are very small for all energies, being less than $\pm 0.1$ and 
usually less than $\pm 0.01$.  The results for each nucleus, ${}^{40}$Ca, 
${}^{89}$Y, ${}^{184}$W, ${}^{197}$Au, and ${}^{238}$U are shown in the bottom 
segment respectively by the solid curve connecting filled circles the 
long-dashed curve connecting filled diamonds, the dashed curve, the solid
curve connecting opaque diamonds, and the dot-dashed curve. 
Of particular note is that the differences between these fit and function
values of the $l_0(E)$ mirror those of the total cross section differences
shown in the top segment, both in energy and with different mass.
It is most likely then that the function form Eq.~(\ref{fnofE}) is a first
order guess and may be improved to meet the reasonable fit criterion we have
set.  That is the subject of ongoing study in which  many more targets 
and more numerous values of energy in the region to 100 MeV are to be used.

\subsection{The parameter $l_0$ as a function of mass}

\begin{figure}
\centering \scalebox{0.8}{\includegraphics{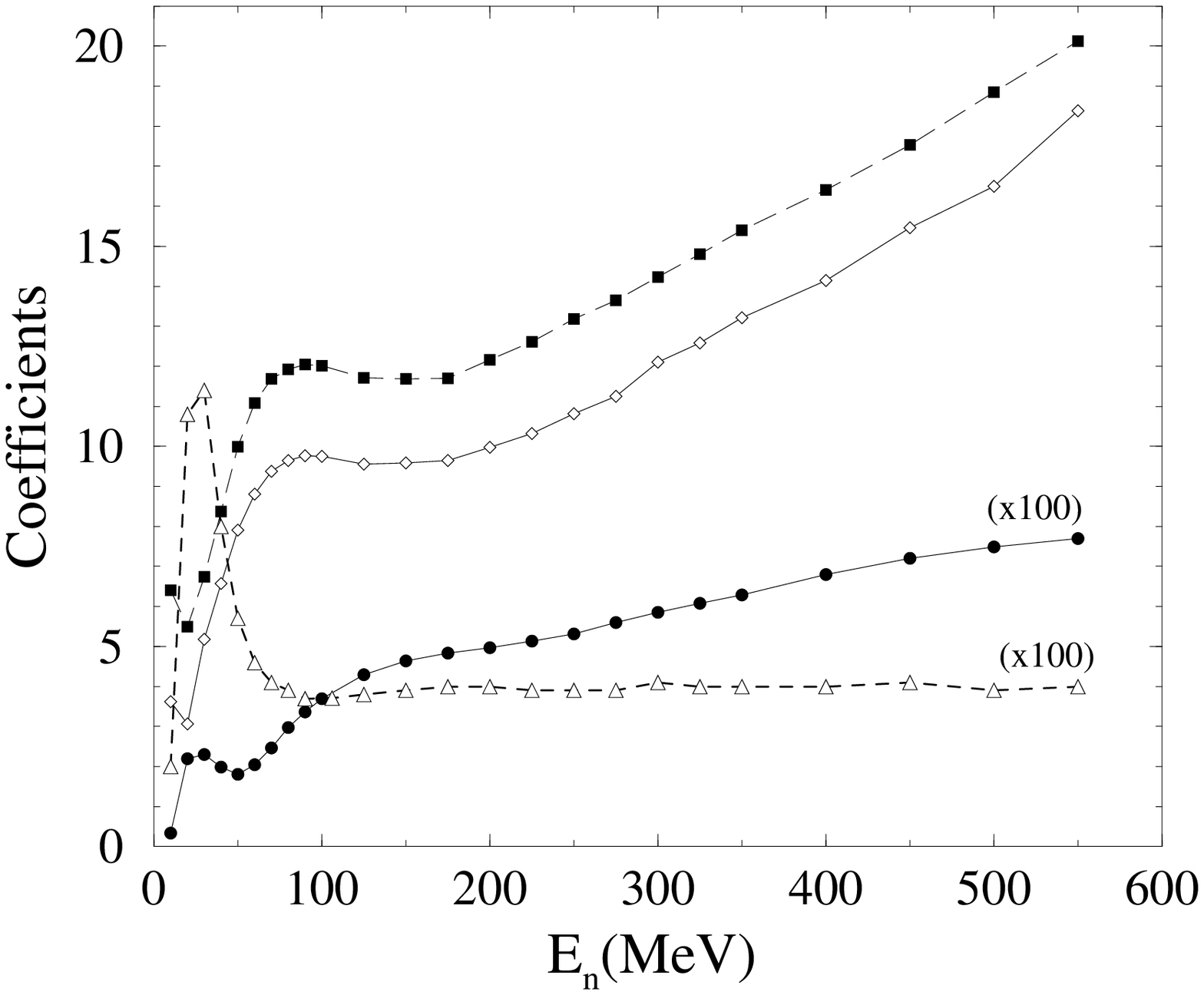}}
\caption{\label{Aparam}
Parameter values of $l_0(A)$ that give best fits to total cross section data.
Details are specified in the text.}
\end{figure}

 As noted above, the $l_0$ parameter values vary smoothly with mass. In fact 
we find that a good representation of those values is given by
\begin{equation}
l_0^{\rm th}(A) = d_1 A + d_2 - d_3 e^{-d_4 A}\ .
\label{FnofA}
\end{equation}
With this mass variation form, the coefficients $d_i$ are as set out in
Table~\ref{FnAparams}.
\begin{table}
\begin{ruledtabular}
\caption{\label{FnAparams}  Values of parameters defining $l_0(A)$}
\begin{tabular}{ccccc}
Energy & $d_1$ & $d_2$ & $d_3$ & $d_4$\\  
\hline
\ 10 & 0.0034 & 6.4 & 3.62 & 0.020 \\
\ 20 & 0.0224  & 5.51 & 3.06 & 0.108 \\
\ 30 & 0.023   & 6.74 & 5.18 & 0.114 \\
\ 40 & 0.0198 & 8.37 & 6.57 & 0.080 \\
\ 50 & 0.018 & 9.99 & 7.91 & 0.057 \\ 
\ 60 & 0.0205 & 11.08 & 8.81 & 0.046 \\
\ 70 & 0.0247 & 11.68 & 9.38 & 0.041 \\ 
\ 80 & 0.0297 & 11.92 & 9.64 & 0.039 \\
\ 90 & 0.0337 & 12.05 & 9.77 & 0.037 \\
 100 & 0.037 & 12.02 & 9.75 & 0.037 \\
 125 & 0.043 & 11.71 & 9.56 & 0.038 \\
 150 & 0.0464 & 11.68 & 9.58 & 0.039 \\
 175 & 0.0484 & 11.70 & 9.64 & 0.040 \\
 200 & 0.0497 & 12.17 & 9.97 & 0.04 \\
 225 & 0.0514 & 12.62 & 10.32 & 0.039 \\
 250 & 0.0532 & 13.19 & 10.82 & 0.039 \\
 275 & 0.056 & 13.65 & 11.25 & 0.039 \\
 300 & 0.0585 & 14.23 & 12.11 & 0.041 \\
 325 & 0.0607 & 14.81 & 12.59 & 0.040 \\
 350 & 0.0628 & 15.41 & 13.22 & 0.040 \\
 400 & 0.068 & 16.40 & 14.14 & 0.040 \\ 
 450 & 0.072 & 17.53 & 15.47 & 0.041 \\
 500 & 0.0748 & 18.85 & 16.50 & 0.039 \\
 550 & 0.077 & 20.13 & 18.39 &  0.040 \\   
\end{tabular}
\end{ruledtabular}
\end{table}
That mass equation with those tabled values of the coefficients gave the nine 
values of $l_0$ for each energy that are connected by a spline curve in 
Fig.~\ref{l0vsA}. The optimal values for these parameters (listed) are shown 
by the diverse set of open and closed symbols. The coefficients defining 
$l_0^{\rm th}(A)$ are portrayed by the various symbols in Fig.~\ref{Aparam}.
Specifically the coefficients are shown by the filled circles ($d_1$), by the
filled squares connected by the long-dashed lines ($d_2$), by the opaque 
diamonds ($d_3$), and by the opaque up-triangles connected by dashed
lines ($d_4$). Again for clarity the actual values found for $d_1$ and $d_4$ 
have been multiplied by a scaling factor. This time that factor is 100.
These mass formula coefficients vary smoothly with energy and one might look
for a convenient function of energy to describe them as well. However, as
we noted earlier with the energy function representation, the choice of this
mass equation resulted solely from inspection of the diagram and so alternate
formulas are not excluded. Therefore it was not sensible to seek a function 
form for the coefficients themselves. In any event, one needs results from
a much larger range of nuclei to study further such mass variations.

\section{Conclusions}

We have found that a simple function of 3 parameters suffice to fit observed
neutron total scattering cross sections from diverse nuclei, ${}^6$Li to 
${}^{238}$U, and for energies ranging from 10 to 600 MeV. That function was 
predicated upon the values of partial total cross sections evaluated using a 
$g$-folding optical potential for scattering. The patterns of the calculated
partial cross sections suggested that two of the parameters, $a$ and 
$\epsilon$, could be set by parabolic functions of energy for all masses. Then
allowing the third parameter $l_0$ to vary, values could be found with which 
the appropriate sum over partial cross sections given by the function form 
exactly match  measured data.  The optimal values of $l_0$ varied smoothly 
with both energy and target mass. 
The energy variations $l_0(E)$ could be 
characterized by yet another simple function form as could the mass variations
$l_0^{\rm th}(A)$.  However, the reasonable fit criterion that final results
remain within $\pm$ 5\% of observation showed that refinement of the 
functional dependencies of the parameter $l_0$ in particular awaits results 
of a far more complete study involving as many 
target masses as possible and for many more energies, particularly below 100 
MeV where the total cross-section data show large scale oscillatory structure.

Nonetheless on the basis of the limited set of nuclei and energies considered,
there is a three parameter function form for partial total cross sections 
that will give neutron total cross sections as required in any application
without recourse to phenomenological optical potential parameter searches.  
One may use tabulations of $l_0(E)$ and interpolations on that table,
or indeed a better data base formed by considering many more energies and many 
more nuclear targets, to
get cross sections satisfying the reasonable fit criterion. A caveat being
that any special gross nuclear structure effect, such as a halo matter
distribution for example, must be separately considered.


\begin{acknowledgments}
We are grateful to Dr. Steven Karataglidis for many useful discussions had,
as well as critical comments made, during the preparation of this manuscript.
This research was supported by a research grant from the Australian Research 
Council and also by the National Science Foundation under Grant No. 0098645.
\end{acknowledgments}

\bibliography{FN-total}

\begin{thebibliography}{11}
\expandafter\ifx\csname natexlab\endcsname\relax\def\natexlab#1{#1}\fi
\expandafter\ifx\csname bibnamefont\endcsname\relax
  \def\bibnamefont#1{#1}\fi
\expandafter\ifx\csname bibfnamefont\endcsname\relax
  \def\bibfnamefont#1{#1}\fi
\expandafter\ifx\csname citenamefont\endcsname\relax
  \def\citenamefont#1{#1}\fi
\expandafter\ifx\csname url\endcsname\relax
  \def\url#1{\texttt{#1}}\fi
\expandafter\ifx\csname urlprefix\endcsname\relax\def\urlprefix{URL }\fi
\providecommand{\bibinfo}[2]{#2}
\providecommand{\eprint}[2][]{\url{#2}}

\bibitem[{\citenamefont{Amos and Deb}(2002)}]{Am02}
\bibinfo{author}{\bibfnamefont{K.}~\bibnamefont{Amos}} \bibnamefont{and}
  \bibinfo{author}{\bibfnamefont{P.~K.} \bibnamefont{Deb}},
  \bibinfo{journal}{Phys. Rev. C} \textbf{\bibinfo{volume}{66}},
  \bibinfo{pages}{024604} (\bibinfo{year}{2002}).

\bibitem[{\citenamefont{Deb and Amos}(2003)}]{De03}
\bibinfo{author}{\bibfnamefont{P.~K.} \bibnamefont{Deb}} \bibnamefont{and}
  \bibinfo{author}{\bibfnamefont{K.}~\bibnamefont{Amos}},
  \bibinfo{journal}{Phys. Rev. C} \textbf{\bibinfo{volume}{67}},
  \bibinfo{pages}{039306} (\bibinfo{year}{2003}).

\bibitem[{\citenamefont{Koning and Delaroche}(2003)}]{Ko03}
\bibinfo{author}{\bibfnamefont{A.~J.} \bibnamefont{Koning}} \bibnamefont{and}
  \bibinfo{author}{\bibfnamefont{J.~P.} \bibnamefont{Delaroche}},
  \bibinfo{journal}{Nucl. Phys. A} \textbf{\bibinfo{volume}{713}},
  \bibinfo{pages}{231} (\bibinfo{year}{2003}).

\bibitem[{\citenamefont{Amos et~al.}(2002)\citenamefont{Amos, Karataglidis, and
  Deb}}]{Amos02}
\bibinfo{author}{\bibfnamefont{K.}~\bibnamefont{Amos}},
  \bibinfo{author}{\bibfnamefont{S.}~\bibnamefont{Karataglidis}},
  \bibnamefont{and} \bibinfo{author}{\bibfnamefont{P.~K.} \bibnamefont{Deb}},
  \bibinfo{journal}{Phys. Rev. C} \textbf{\bibinfo{volume}{65}},
  \bibinfo{pages}{064618} (\bibinfo{year}{2002}).

\bibitem[{\citenamefont{Amos et~al.}(2000)\citenamefont{Amos, Dortmans, von
  Geramb, Karataglidis, and Raynal}}]{Am00}
\bibinfo{author}{\bibfnamefont{K.}~\bibnamefont{Amos}},
  \bibinfo{author}{\bibfnamefont{P.~J.} \bibnamefont{Dortmans}},
  \bibinfo{author}{\bibfnamefont{H.~V.} \bibnamefont{von Geramb}},
  \bibinfo{author}{\bibfnamefont{S.}~\bibnamefont{Karataglidis}},
  \bibnamefont{and} \bibinfo{author}{\bibfnamefont{J.}~\bibnamefont{Raynal}},
  \bibinfo{journal}{Adv. in Nucl. Phys.} \textbf{\bibinfo{volume}{25}},
  \bibinfo{pages}{275} (\bibinfo{year}{2000}).

\bibitem[{\citenamefont{von Geramb et~al.}(1998)\citenamefont{von Geramb, Amos,
  Labes, and Sander}}]{Ge98}
\bibinfo{author}{\bibfnamefont{H.}~\bibnamefont{von Geramb}},
  \bibinfo{author}{\bibfnamefont{K.}~\bibnamefont{Amos}},
  \bibinfo{author}{\bibfnamefont{H.}~\bibnamefont{Labes}}, \bibnamefont{and}
  \bibinfo{author}{\bibfnamefont{M.}~\bibnamefont{Sander}},
  \bibinfo{journal}{Phys. Rev. C} \textbf{\bibinfo{volume}{58}},
  \bibinfo{pages}{2249} (\bibinfo{year}{1998}).

\bibitem[{\citenamefont{Lagoyannis et~al.}(2001)}]{La01}
\bibinfo{author}{\bibfnamefont{A.}~\bibnamefont{Lagoyannis}}
  \bibnamefont{et~al.}, \bibinfo{journal}{Phys. Lett.}
  \textbf{\bibinfo{volume}{B518}}, \bibinfo{pages}{27} (\bibinfo{year}{2001}).

\bibitem[{\citenamefont{Abfalterer et~al.}(2001)\citenamefont{Abfalterer,
  Bateman, Dietrich, Finlay, Haight, and Morgan}}]{Abf01}
\bibinfo{author}{\bibfnamefont{W.~P.} \bibnamefont{Abfalterer}},
  \bibinfo{author}{\bibfnamefont{F.~B.} \bibnamefont{Bateman}},
  \bibinfo{author}{\bibfnamefont{F.~S.} \bibnamefont{Dietrich}},
  \bibinfo{author}{\bibfnamefont{R.~W.} \bibnamefont{Finlay}},
  \bibinfo{author}{\bibfnamefont{R.~C.} \bibnamefont{Haight}},
  \bibnamefont{and} \bibinfo{author}{\bibfnamefont{G.~L.}
  \bibnamefont{Morgan}}, \bibinfo{journal}{Phys. Rev. C}
  \textbf{\bibinfo{volume}{63}}, \bibinfo{pages}{044608}
  (\bibinfo{year}{2001}).

\bibitem[{\citenamefont{Finlay et~al.}(1993)\citenamefont{Finlay, Abfalterer,
  Fink, Montei, Adami, Lisowski, Morgan, and Haight}}]{Fin93}
\bibinfo{author}{\bibfnamefont{R.~W.} \bibnamefont{Finlay}},
  \bibinfo{author}{\bibfnamefont{W.~P.} \bibnamefont{Abfalterer}},
  \bibinfo{author}{\bibfnamefont{G.}~\bibnamefont{Fink}},
  \bibinfo{author}{\bibfnamefont{E.}~\bibnamefont{Montei}},
  \bibinfo{author}{\bibfnamefont{T.}~\bibnamefont{Adami}},
  \bibinfo{author}{\bibfnamefont{P.~W.} \bibnamefont{Lisowski}},
  \bibinfo{author}{\bibfnamefont{G.~L.} \bibnamefont{Morgan}},
  \bibnamefont{and} \bibinfo{author}{\bibfnamefont{R.~C.}
  \bibnamefont{Haight}}, \bibinfo{journal}{Phys. Rev. C}
  \textbf{\bibinfo{volume}{47}}, \bibinfo{pages}{237} (\bibinfo{year}{1993}).

\bibitem[{\citenamefont{Brown}(2000)}]{Br00}
\bibinfo{author}{\bibfnamefont{B.~A.} \bibnamefont{Brown}},
  \bibinfo{journal}{Phys. Rev. Lett.} \textbf{\bibinfo{volume}{85}},
  \bibinfo{pages}{5296} (\bibinfo{year}{2000}).

\bibitem[{\citenamefont{Karataglidis et~al.}(2002)\citenamefont{Karataglidis,
  Amos, Brown, and Deb}}]{Ka02}
\bibinfo{author}{\bibfnamefont{S.}~\bibnamefont{Karataglidis}},
  \bibinfo{author}{\bibfnamefont{K.}~\bibnamefont{Amos}},
  \bibinfo{author}{\bibfnamefont{B.~A.} \bibnamefont{Brown}}, \bibnamefont{and}
  \bibinfo{author}{\bibfnamefont{P.~K.} \bibnamefont{Deb}},
  \bibinfo{journal}{Phys. Rev. C} \textbf{\bibinfo{volume}{65}},
  \bibinfo{pages}{044306} (\bibinfo{year}{2002}).

\end{thebibliography}

\end{document}